# Vitality of Neural Networks under Reoccurring Catastrophic Failures

Shira Sardi[1,†], Amir Goldental[1,†], Hamutal Amir[2], Roni Vardi[2] and Ido Kanter[1,2,*]

[1]Department of Physics, Bar-Ilan University, Ramat-Gan 52900, Israel

[2]Gonda Interdisciplinary Brain Research Center and the Goodman Faculty of Life Sciences, Bar-Ilan University, Ramat-Gan 52900, Israel

†These authors contributed equally to this work.

*Corresponding author: ido.kanter@biu.ac.il

**Catastrophic failures are complete and sudden collapses in the activity of large networks such as economics, electrical power grids and computer networks, which typically require a manual recovery process. Here we experimentally show that excitatory neural networks are governed by a non-Poissonian reoccurrence of catastrophic failures, where their repetition time follows a multimodal distribution characterized by a few tenths of a second and tens of seconds timescales. The mechanism underlying the termination and reappearance of network activity is quantitatively shown here to be associated with nodal time-dependent features, neuronal plasticity, where hyperactive nodes damage the response capability of their neighbors. It presents a complementary mechanism for the emergence of Poissonian catastrophic failures from damage conductivity. The effect that hyperactive nodes degenerate their neighbors represents a type of local competition which is a common feature in the dynamics of real-world complex networks, whereas their spontaneous recoveries represent a vitality which enhances reliable functionality.**

## Introduction

Catastrophic failures[1-4] in the activity of a network[5-12] may occur as a result of a cascading failure[4,13,14], in which the failure of one node can trigger the failure of other connected nodes in a chain reaction. The number of failing nodes rapidly increases until the activity of the entire network runs into an irrecoverable collapse. A recovery typically requires an intensive external action and interruption, such as the replacement of some of the failing parts and a reset or re-synchronization of the entire system. Consequently, a unique event of total collapse prevents the continuation of the autonomous activity of the interconnected system, although it was recently shown that in case that the failed parts recover spontaneously the network itself can recover and fail repeatedly[10].

Neural networks that exhibit catastrophic failures lead to a silence of activity which results in the loss of computational capabilities. Therefore, the consistent functionality of the brain has to include either a mechanism which practically eliminates the probability of such catastrophic failures[15,16], e.g. strokes, or a robust biological mechanism which recovers the network from such synchronized failures. Here we experimentally show that indeed the second mechanism is realized in the activity of neural networks. The mechanism for the reoccurrences of total collapses is neuronal plasticity in the form of neuronal response failures, which dynamically emerge in an overshoot manner. Surprisingly, the same mechanism, the neuronal plasticity, is also responsible for the self-recovery mechanism from these total collapses.

The experimental setup consists of cortical tissue culture of ~4 cm$^2$ size (Fig. 1a), with a multi-electrode array in the center of the tissue (Fig. 1a, Online Methods). The multi-electrode array consists of 60 extra-cellular electrodes, separated by 0.5 mm, and is responsible for sampling the spontaneous firing activity of the neural network, consisting of around one million interconnected neurons[17] (Fig. 1b-c, Online Methods). Results are presented for excitatory networks (Online Methods), however, the main conclusions remain valid also for networks consisting of a mixture of excitatory and inhibitory connections (Supplementary Fig. S1).

**Results**

The raster plot of the activity recorded by the 60 electrodes over a period of one hour is exemplified in a snapshot of 150 seconds (Fig. 1b). The activity is governed by macroscopic cooperation among neurons comprising the network, in the form of burst activities[18-22] (Fig. 1b,c), separated by periods of at least 30 milliseconds of vanishing activity (Online Methods). The duration of a burst is typically a few dozens of milliseconds and can be extended to several hundreds of milliseconds. The visible peaks in the autocorrelation of the network's firing rate, few dozens of Hertz (Fig. 1d), are neither sharp nor isolated and are surrounded by background noise as a result of fluctuations in the structure of different bursts (Supplementary Fig. S4). These oscillations[23] in the network activity stem from neuronal plasticity and were explained both by simulations and by an analytical description[23]. However, the mechanism underlying the long time-lags between bursts (Fig. 1b) and their statistics were not fully explained yet and are at the center of this study.

The statistics of the time-lags between consecutive network bursts (Online Methods), silent periods, consist of a multimodal distribution (Fig. 1e). The short time-lags, S, range from several tens to a few hundreds of milliseconds, whereas the long time-lags, L, range from several to tens of seconds, and a vanishing fraction of events occurs at ~[0.5, 1.5] seconds. A correlation between consecutive time-lags was examined using the following two statistical measurements. The first measurement is the probability for the occurrence of two long silent periods (L) separated by m short silent periods (S), which was found to be in a good agreement with a Poisson process (Fig. 1f). The second measurement is the probabilities for the 8 possible combinations of 3 consecutive silent periods (Fig. 1g). Both statistical measurements strongly indicate that silent periods are sampled independently from the multimodal distribution (Fig. 1e).

We now turn to show that the time-lags between bursts are controlled by the time-dependent features of the neurons (nodes), neuronal plasticity[24], as opposed to synaptic (link) plasticity. The recorded firing rate of the neurons during a burst may reach several hundreds of Hertz (Fig. 2a and Supplementary Figs. S2 and S3), and the inter-spike-intervals, ISIs, practically vanish below ~2 milliseconds, representing the typical duration of the neuronal absolute refractory period. Since neurons fire at very high frequencies during bursts (Fig. 2a and Fig. 1b-c), a neuron in a highly connected network is most likely continuously and strongly (supra-threshold) stimulated. Consequently, the ISI probability density function is similar to an exponential decay function, shifted by the absolute refractory period.

When a neuron is stimulated at high frequency, it goes through a transient between two phases, as reflected by the neuronal response probability and by the neuronal response latency[24], NRL, which measures the time-lag between a stimulation and its corresponding evoked spike. At the initial phase, the neuron's firing rate is equal to its stimulation rate, the neuron has no response failures (Fig. 2b) and its NRL gradually increases (Fig. 2c). As the stimulation period goes on, the neuron enters the intermittent phase, where the firing frequency and the NRL are saturated (Fig. 2b). This neuronal maximal firing frequency, $f_c$, is controlled by stochastic neuronal response failures (Fig. 2c), and varies among neurons, typically in the range of [1, 30] Hz. The time scale $1/f_c$ is the source for the revival of the bursts every several dozens or hundreds of milliseconds[25].

The source for a much slower cooperative behavior timescale, tens of seconds, between bursts (Fig. 1e) is a consequence of the crossover between the two abovementioned response phases of each neuron. For many neurons, the crossover is accompanied by an overshoot behavior, where the NRL increases above the saturated NRL at the intermittent phase, and the response probability drops substantially to an almost vanishing firing frequency before increasing again to $f_c$ (Fig. 2c). These several seconds of overshoot represent a type of a "kick down" mechanism which accelerates the accumulated averaged firing frequency faster towards $f_c$ (dashed curve, Fig. 2c). In case that a sufficient fraction of neurons are in the overshoot region, the ignition of a burst, as a collective behavior, is blocked for many seconds. This is the origin for the long inter-burst-interval (Fig. 1e). We expect this period to extend to tens of seconds, since periods of overshoot are not fully synchronized among neurons. This explanation assumes that the neuron is continuously stimulated; however, one cannot exclude the possibility that after a short period without stimulations, as in the case of inter-burst-intervals, the NRL decays towards the initial NRL. In such a scenario, the neuron would retreat from the intermittent phase back to the initial phase which is characterized by a high response probability.

For random neural networks, the average response probability can be integrated to a simplified toy map, describing the dynamics at the beginning of a burst:

$$R_t = K \cdot p_s(t) \cdot R_{t-\Delta} \qquad (1)$$

where $R_t$ is the fraction of firing neurons at time t, $\Delta$ is the average time delay between neurons, $p_s(t)$ is the response probability of a neuron averaged over the network and K stands for the average neuronal connectivity and represents the momentary gain of the network firing rate per unit time, $\Delta$. In case that there are no response failures, the gain is expected to be higher than 1, otherwise the activity consists of solely local avalanches[26,27]. Only when the condition

$$p_s(t) > k^{-1} \qquad (2)$$

is achieved, a cooperative burst composed of most of the network can start evolving, similar to the achievement of site percolation threshold[28-31]. After a sequence of nearby bursts separated by short silent periods (S) ends, $p_s$ is very low as a result of the high activity of the network (Fig. 2a) and it starts to increase with time as more neurons are fading out of their intermittent phase. The recovery time of $p_s(t)$ (Fig. 1e) is associated with the long inter-burst-intervals, eq. (2), and is examined and estimated experimentally.

Qualitatively, after a long IBI, e.g. ten seconds, many of the neurons decay to their initial latency and their response probability increases towards unity. As bursts evolve, neurons are stimulated and fire at high frequencies (Fig. 2a) and are driven to the overshoot phase and beyond to the intermittent phase (Fig. 2c), resulting in a decrease of $p_s$. During the intermittent phase the response probability of a neuron is inversely proportional to the stimulation frequency[24], $f_c/f$, and almost vanishes during a burst (Fig. 2a).

To estimate the recovery time of $p_s(t)$ of a single neuron we define the following appropriate stimulations scheduling. A long silence period (L) occurs on the average after several bursts separated by short silence periods (S) (Fig. 1e and Supplementary Fig. S1), where in each burst a neuron fires few dozens of times (Fig. 2a). Accordingly, we define a single neuron experiment with the following stimulations scheduling. A "bunch", imitating a burst, is a set of 26 stimulations, at 130 Hz. The neuron is stimulated by 5 bunches with 200 milliseconds of silence between them (Fig. 3 upper panels), which are repeated after a relatively long time-lag in the range of [1, 15] seconds. This structure of stimulations, few dense bunches separated by long time-lags, imitates the activity of the network (Fig. 1e). It is evident that the response probability of a single neuron decreases as more bunches are given (Fig. 3). This trend is consistent with the recorded activity of the network, where the probability for spike detection is higher at the beginning of the burst (Zoon-in Fig. 1C and supplementary Fig. S4). The response probability for the first bunch (green dots in Fig. 3) and the last bunch among the five (purple dots in Fig. 3) was estimated as a function of the long time-lags. Results clearly indicate that after a time-lag of ~10 seconds without stimulations the neuronal response probability recovers (Fig. 3 bottom panel). This timescale of 10 seconds is associated with the period necessary for a neuron to pullout from the intermittent phase and is the main mechanism which dictates the origination of the next burst. As expected, results also indicate that the response probability of the last bunch is much lower compared to other bunches and in particular in comparison to the first bunch.

## Discussion

We present experimental data where conductance failures of a node in a large neural network result from an overload, hyperactivity of neighboring nodes. This mechanism is opposed to the damage conductivity paradigm[10] where the damaged nodes degenerate their neighboring nodes. On the macroscopic level, these two mechanisms lead to different kinds of dynamics. Specifically, in the presented results the transition between the two phases of the network, active and "dead", is non-Poissonian (multimodal distribution, Fig. 1e), but has characteristic timescales. These timescales result from the memory of nodes which leads to a non-Markovian process and are expected to be independent of the size of the network (eq. (2)). On the other hand, the damage conductivity paradigm leads to Poissonian statistics where transitions strongly depend on the size of the network.

The variability among the structure of bursts and the distribution of the IBIs contains information on the structure of the network and might help to infer the network topology[32-36]. For example, under the assumption of a random network, the average effective degree per node might be inferred by measuring the average response probability of neurons during the network dynamics (see eq. 2). However, the possibility to infer the detailed topology of a general network from its dynamical activity is a challenge.

Finally, the perceptual significance of bursts on learning and cognition processes in neural networks is unclear and it might function as a limited reset mechanism. It hints on the usefulness of stochastic elements which their current activity depends on network's activity history. It is then expected that similar types of nodal plasticity might generate more immune and robust networks in various realizations including power transmission, computer networking and electrical grids, indicating the advantage of elements with conductance failures.

## Methods

**Animals.** All procedures were in accordance with the National Institutes of Health Guide for the Care and Use of Laboratory Animals and Bar-Ilan University Guidelines for the Use and Care of Laboratory Animals in Research and were approved and supervised by the Institutional Animal Care and Use Committee.

**Culture preparation.** Cortical neurons were obtained from newborn rats (Sprague-Dawley) within 48 h after birth using mechanical and enzymatic procedures. The cortical tissue was digested enzymatically with 0.05% trypsin solution in phosphate-buffered saline (Dulbecco's PBS) free of calcium and magnesium, and supplemented with 20 mM glucose, at 37°C. Enzyme treatment was terminated using heat-inactivated horse serum, and cells were then mechanically dissociated. The neurons were plated directly onto substrate-integrated multi-electrode arrays (MEAs) and allowed to develop functionally and structurally mature networks over a time period of 2-3 weeks in vitro, prior to the experiments. Variability in the number of cultured days in this range had no effect on the observed results. The number of plated neurons in a typical network was in the order of 1,300,000, covering an area of about ~4 cm$^2$. The preparations were bathed in minimal essential medium (MEM-Earle, Earle's Salt Base without L-Glutamine) supplemented with heat-inactivated horse serum (5%), glutamine (0.5 mM), glucose (20 mM), and gentamicin (10 g/ml), and maintained in an atmosphere of 37°C, 5% $CO_2$ and 95% air in an incubator as well as during the electrophysiological measurements.

**Synaptic blockers.** Spontaneous network activity recordings were conducted on cultured cortical neurons in which inhibition was reduced by a pharmacological block of GABAergic synapses. For each culture 2 µl of 5 µM Bicuculline were used.
Single neuron stimulation and recording experiments were conducted on cultured cortical neurons that were functionally isolated from their network by a pharmacological block of glutamatergic and GABAergic synapses. For each culture 20 µl of a cocktail of synaptic blockers were used, consisting of 10 µM CNQX (6-cyano-7-nitroquinoxaline-2,3-dione), 80 µM APV (amino-5-phosphonovaleric acid) and 5 µM Bicuculline. This cocktail did not block the spontaneous network activity completely, but rather made it sparse. At least one hour was allowed for stabilization of the effect.

**Stimulation and recording.** An array of 60 Ti/Au/TiN extracellular electrodes, 30 µm in diameter, and spaced 500 µm from each other (Multi-Channel Systems, Reutlingen, Germany) was used. The insulation layer (silicon nitride) was pre-treated with polyethyleneimine (0.01% in 0.1 M Borate buffer solution). A commercial setup (MEA2100-2x60-headstage, MEA2100-interface board, MCS, Reutlingen, Germany) for recording and analyzing data from two 60-electrode MEAs was used, with integrated data acquisition from 120 MEA electrodes and 8 additional analog channels, integrated filter amplifier and 3-channel current or voltage stimulus generator (for each 60 electrode array). For the stimulations in the experiment in figure 3 mono-phasic square voltage pulses were used, in the range of [-800, -500] mV and [60, 400] µs. Each channel was sampled at a frequency of 50k samples/s, thus the recorded action potentials and the changes in the neuronal response latency were measured at a resolution of 20 µs.

**Cell selection.** For the single neuron experiment, a neuron was represented by a stimulation source (source electrode) and a target for the stimulation – the recording electrode (target electrode). These electrodes (source and target) were selected as the ones that evoked well-isolated, well-formed spikes and reliable response with a high signal-to-noise ratio. This examination was done with a stimulus intensity of -800 mV and a duration of 200 µs using 25 repetitions at a rate of 5 Hz, followed by 1200 repetitions at a rate of 10 Hz.

**Data analysis.** Analyses were performed in a Matlab environment (MathWorks, Natwick, MA, USA). The reported results were confirmed based on at least eight experiments each, using different sets of neurons and several tissue cultures. For the recordings of spontaneous network activity, the recorded data (voltage) was filtered by convolution with a Gaussian that has a STD of 0.1 ms, where the threshold for action potential detection was defined to be -6 times the STD of this convolution. For the experiment shown in Fig. 3, evoked spikes were detected by threshold crossing using a detection window of 1-10 ms following the beginning of an electrical stimulation.

Bursts were defined using a rate vector. The rate vector is the averaged firing frequency that was detected from all the 60 electrodes of the MEA, over a time windows of 2 ms, i.e.

$$r(t = n \cdot 2ms) = \frac{\int_{t-1ms}^{t+1ms} \sum_{t^*} \delta(t' - t^*) \, dt'}{2ms}$$

where t is the relevant time, n is an integer and the sum is over all spike times, t*, recorded by the MEA. Values of r below 1 spike per ms, i.e. 2 spikes/time window, were set to zero.

A beginning of a burst is identified when r>0 after at least 30 ms of silence (r=0). The end of a burst is defined as a point where r>0 and is followed by a silence of at least 30 ms. During a burst there is no time period larger than 30 ms that is all zeros in the rate vector. Inter-burst interval (IBI) is defined as the duration between an end of a burst and the beginning of the consecutive burst.

# Figures

**Figure 1. Bursts of spontaneous activity in neural cultures. (a1)** A micro-electrode array (MEA), the orange circle with diameter of ~2.2 cm represents the tissue culture area of

~1.3 million cortical neurons (Online Methods). **(a2)** Zoom in on the blue square in a1 showing the arrangement of the 60 electrodes separated by 500 μm. **(a3)** Zoom in on one electrode, showing neurons and connections. **(b)** A raster plot of the network's spontaneous activity, recorded by the 60 electrodes over 150 seconds out of 60 minutes. **(c)** Zoom in of 10 seconds of the gray area in b (left) and 2 seconds out of the 10 seconds (right). The ~400 ms represents a short inter-burst-interval (IBI) (Online Methods). **(d)** Autocorrelation of the rate of the entire recording in b. **(e)** A normalized histogram of the IBIs presented in log scale (Online Methods), a multimodal distribution is observed. Short/long IBIs are denoted by S/L, respectively. **(f)** A log-linear plot of the probability for m consecutive short IBIs bounded by long IBIs as a function of m (green circles), and the geometric distribution $P=P_L \cdot P_S^m$ (gray line) assuming independent events with probabilities $P_S$ and $P_L=1-P_S$ for S and L IBIs, respectively. **(g)** The eight measured probabilities (green circles) for three consecutive IBIs, where S/L stand for short/long IBIs, and the compared probabilities assuming independent events taken from $P_S$ and $P_L$ (gray circles).

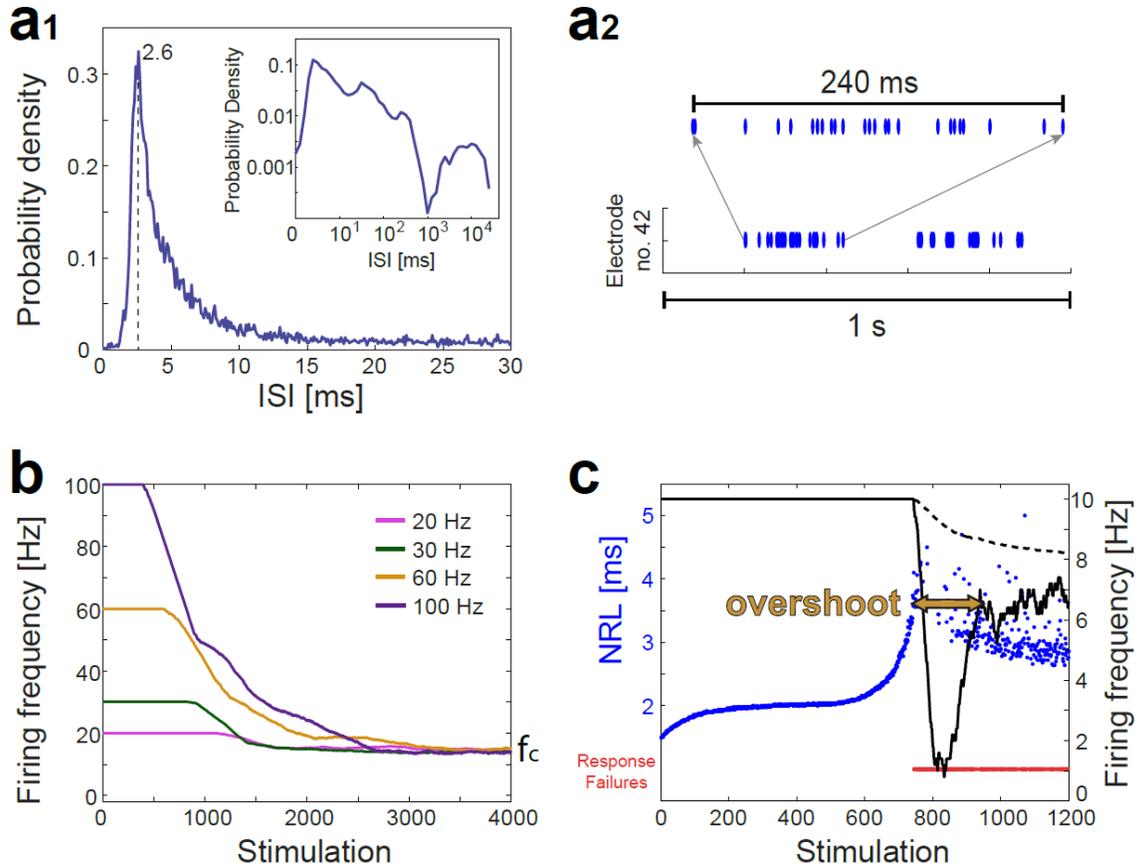

**Figure 2. Activity collapses as a result of the neuronal plasticity. (a1)** The probability density of inter-spike-intervals, ISIs, less than 30 ms, of an electrode from Fig. 1b (Online Methods), and the entire range of the ISIs (inset). **(a2)** Zoom in of one electrode recording. **(b)** The firing frequency of a neuron stimulated at 20, 30, 60 and 100 Hz calculated using sliding windows of 500 stimulations, or the maximal available one for stimulations 1 to 500, indicating a maximal firing frequency, $f_c$~17 Hz, independent of the stimulation frequency. **(c)** Neuronal response latency (NRL) of a neuron stimulated at 10 Hz (blue dots), and its response failures (red dots). The averaged firing frequency calculated using sliding windows of 50 stimulations (black) and the averaged accumulated firing frequency (dashed black line). The overshoot at the transition to the intermittent phase represents a kind of "kick down" mechanism to saturate the firing frequency (fig. 2b).

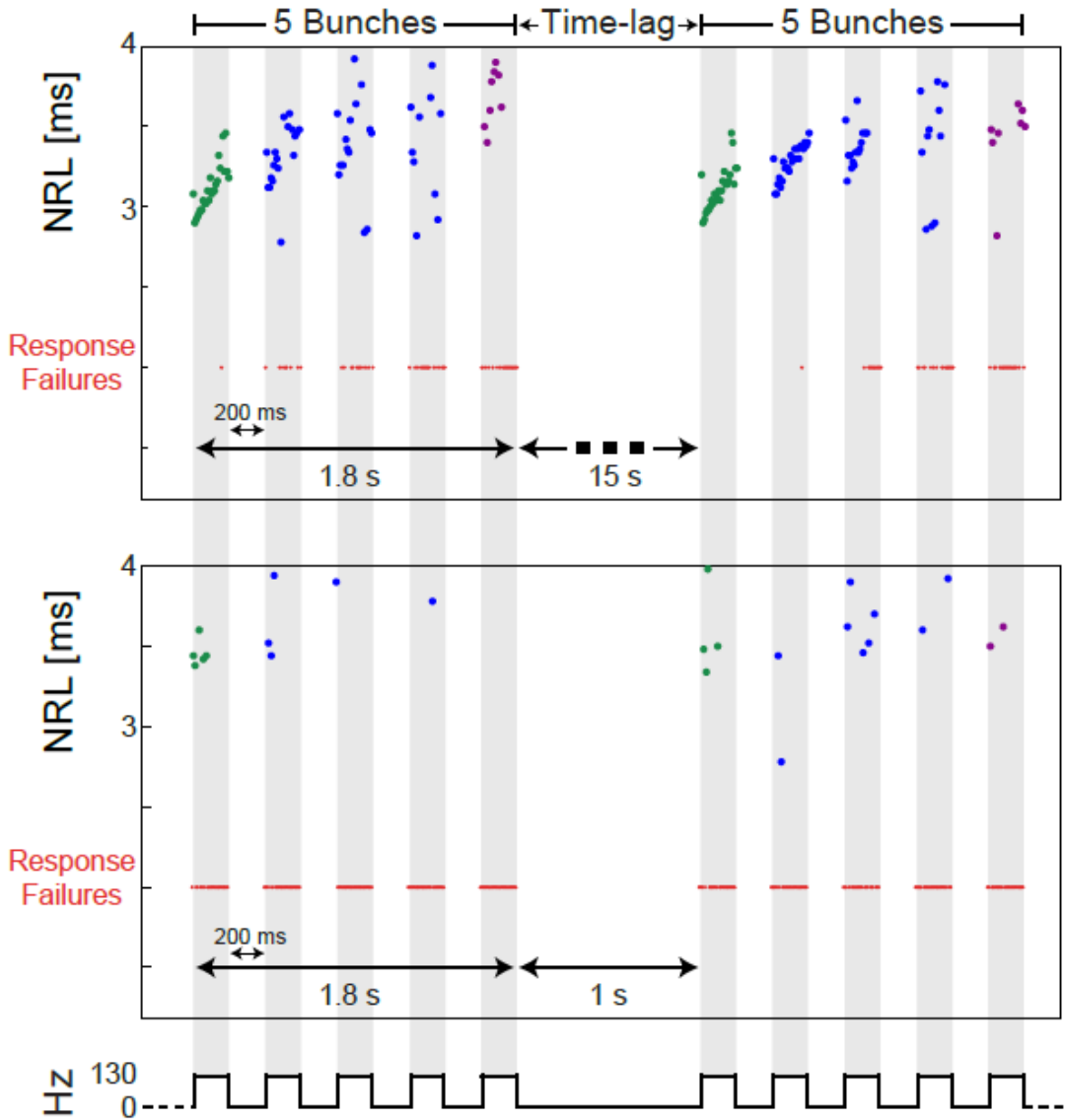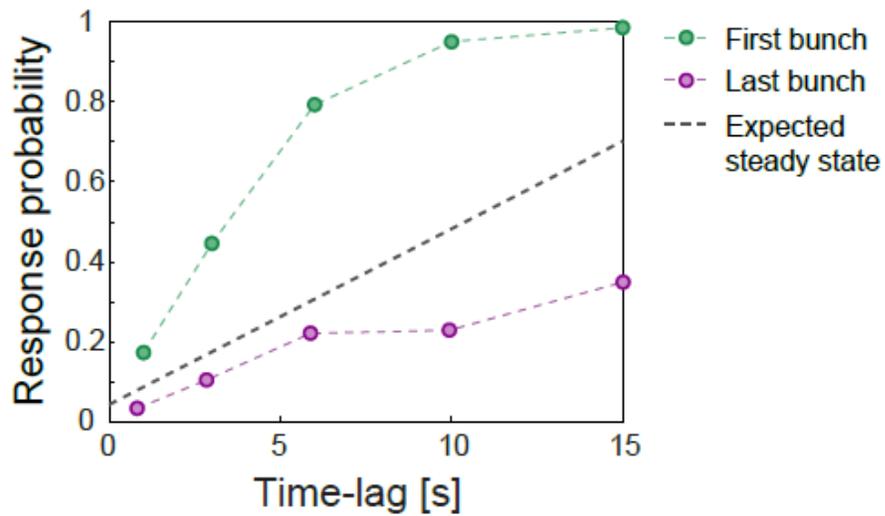

**Figure 3. Reentrance to the intermittent phase control the time-lags between catastrophic failures.** Neuronal response latency (NRL) for repeated stimulation bunches of a neuron. The neuron is stimulated by 5 bunches of 26 stimulations at 130 Hz (gray zones) separated by 200 ms break (the average stimulation frequency is ~72.2 Hz (26x5=130 stimulations in 1.8 s)). These sets of 5 bunches are separated by a longer time-lag. The NRL is denoted for the first bunch (green), last bunch (purple), and for the rest three bunches (blue) as well as the response failures (red). In the upper panel the long time-lag are equal to 15 seconds, and in the middle panel the long time-lag are equal to 1 second. Lower panel: The response probability of the first bunch (green circles) and the last bunch (purple circles) as a function of the long time-lags between bunches. Results are derived from 25 consecutive stimulation bunches. The gray dashed line is the expected average response probability, $f_c/f$, for the stimulated neuron, characterized by $f_c$~5.7 Hz, and f is the equivalent periodic stimulation frequency (=130/[1.8 seconds + time-lag]).

**Author contributions**

SS and RV prepared the cultural tissues and the experimental materials. SS, AG, RV and HA performed the in-vitro experiments; AG and SS analyzed the data; AG developed the theoretical framework. AG and SS wrote the manuscript. I.K. initiated the study and supervised all aspects of the work. All authors discussed the results and commented on the manuscript.

**Acknowledgments**

We thank Moshe Abeles for stimulating discussions. This research is partially supported by the Israel Science foundation.

# Supplementary Information

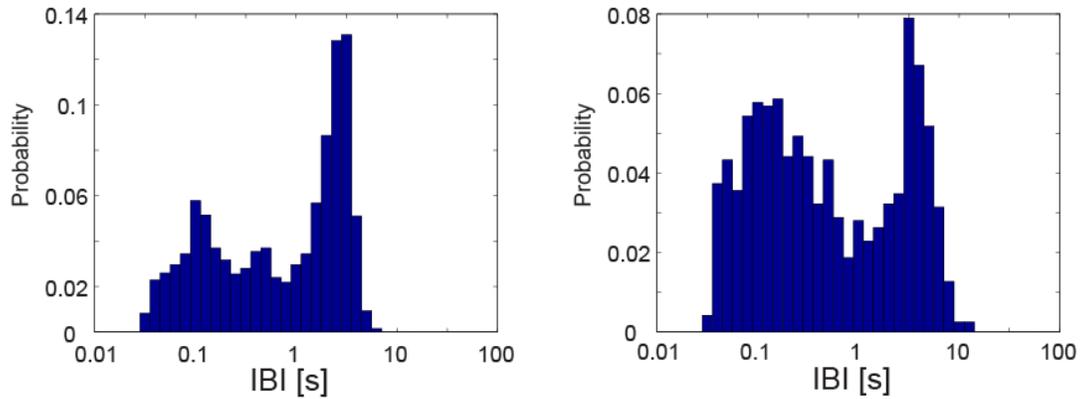

**Figure S1. Inter burst intervals (IBIs) in non-blocked neural cultures.** Two examples of normalized histograms for the IBIs, presented in log-scale (Online Methods), from two different non-blocked neural cultures consisting of excitatory and inhibitory connections (synapses). Results indicate two main maxima in each histogram, around 100 ms and several seconds, however, the bimodal distribution is not as clear as in excitatory networks (Fig. 1).

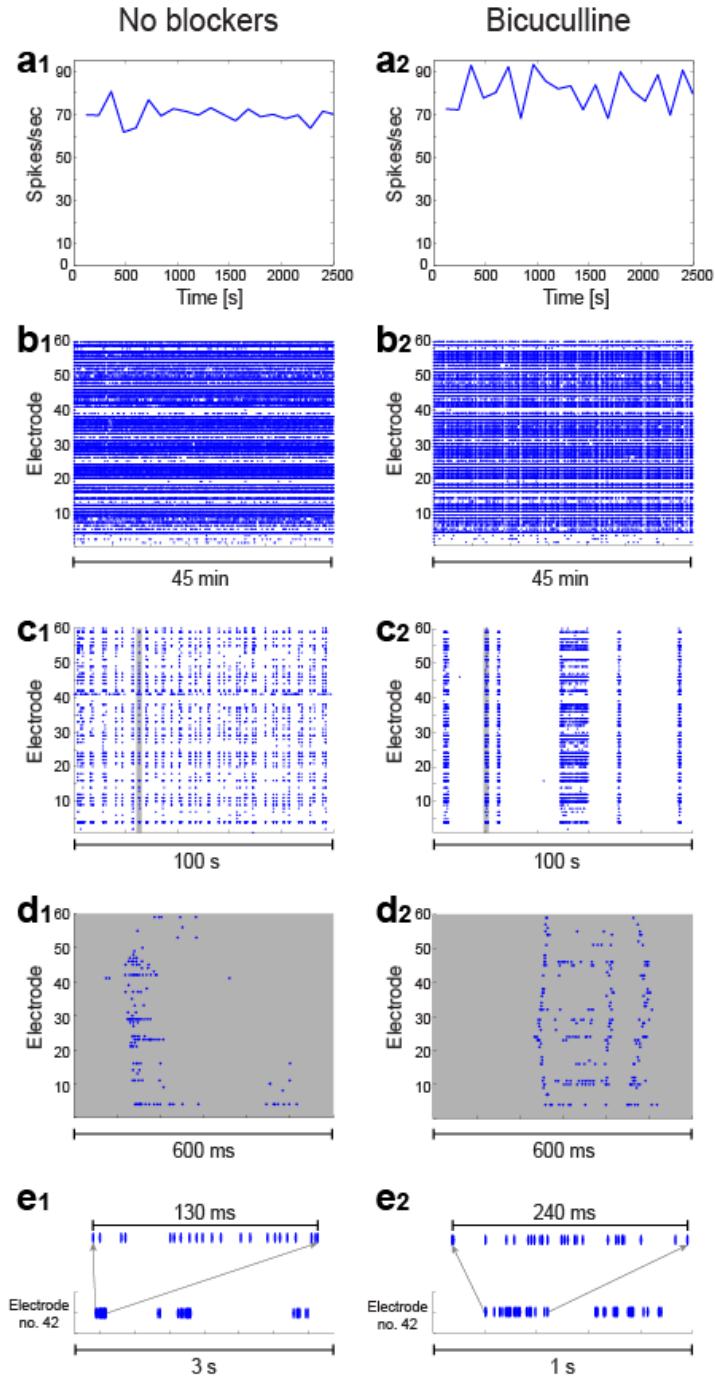

**Figure S2. Spontaneous activity in neural cultures.** Averaged firing rate using sliding window of 2 minutes (Online Methods) for a non-blocked culture **(a1)** and a blocked culture (where Bicuculline was added to block inhibition, Online Methods) **(a2)**. Results indicate that blocking inhibition does not substantially change the level of averaged firing activity, i.e. the average firing rate per electrode is around 1 Hz . A raster plot of the spontaneous activity, recorded from the 60 electrodes over 45 minutes, for a non-blocked culture **(b1)** and a blocked culture **(b2)**. 100 seconds out of 45 minutes in **(b1)** and **(b2)** is presented in **(c1)** and **(c2),** respectively, and a zoom in of 600 ms of the gray area **(d1)** and **(d2)**. Zoom in of one electrode recording in a non-blocked culture **(e1)** and a blocked culture **(e2)**, indicating a few dozens of recorded spikes in a burst.

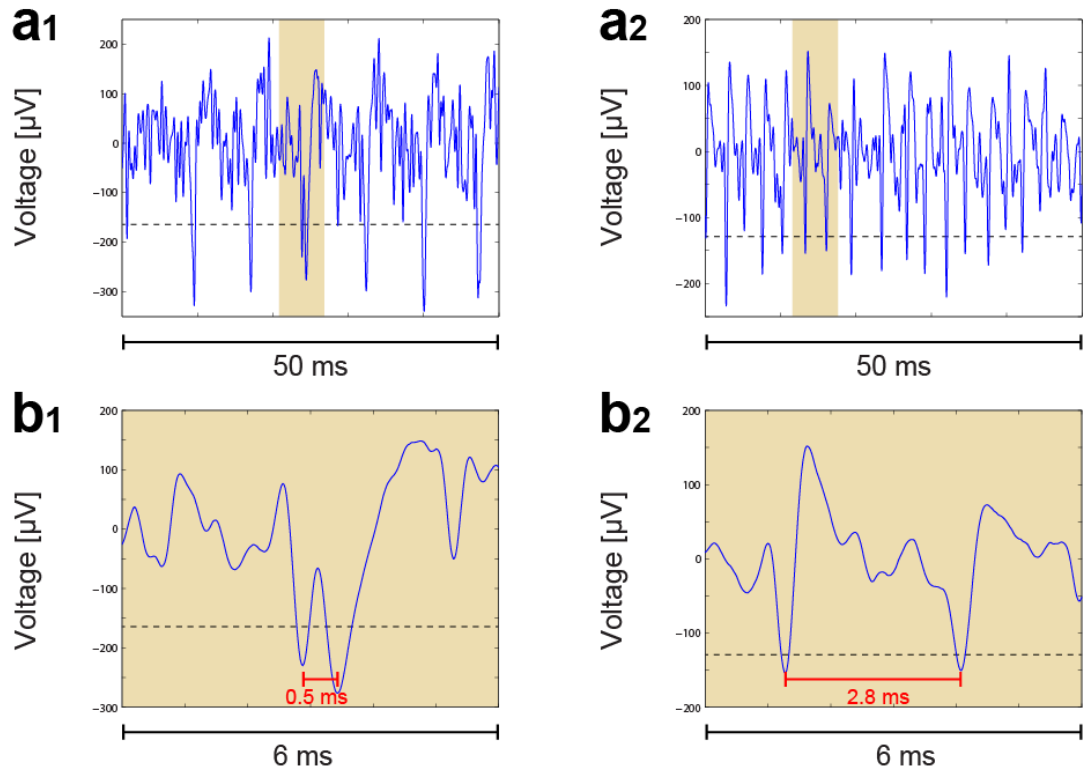

**Figure S3. Voltage recordings.** Recorded voltage of one electrode in a blocked culture (a cocktail of synaptic blockers was added to block excitation and inhibition, Online Methods). The black dashed line presents the threshold for spike detection (Online Methods). An electrode recording from more than one neuron **(a1)** and an electrode recording from a single neuron **(a2)**. Panels **(b1)** and **(b2)** present a zoom-in of the colored areas in **(a1)** and **(a2)**, respectively.

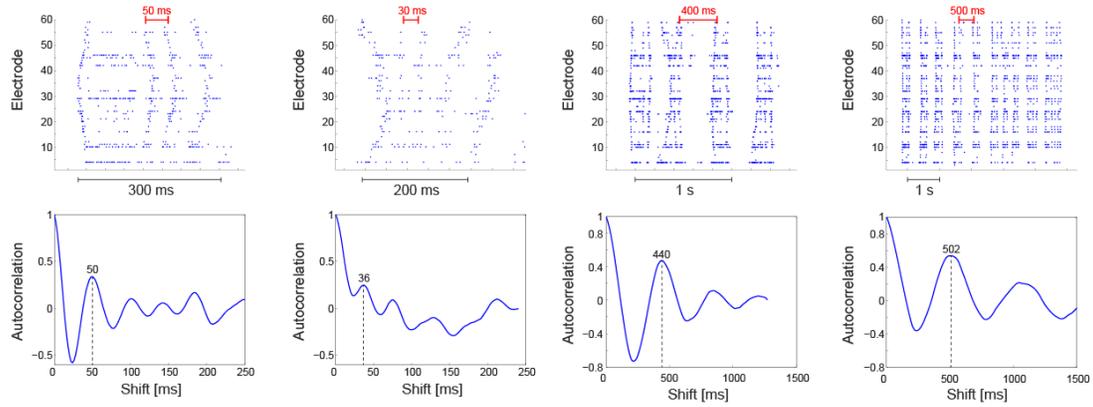

**Figure S4. Autocorrelation of bursts in neural cultures.** Different sets of bursts (taken from the same neural culture) separated only by short IBIs (upper panel) and the autocorrelation on their rate, respectively (lower panel). The visibility of the peaks in the autocorrelation is enhanced in comparison to Fig. 1d in the manuscript.